\documentclass{aastex}
\usepackage{url}

\RequirePackage{color}

\begin{document}

\title{The triple system AT Mic AB + AU Mic in the $\beta$ Pictoris Association}
\shorttitle{The triple system AT Mic AB + AU Mic }
\shortauthors{Messina et al.}

\author{Sergio Messina\altaffilmark{1}, Giuseppe Leto\altaffilmark{1}, Isabella Pagano\altaffilmark{1}} 
\affil{INAF- Catania Astrophysical Observatory, via S.Sofia, 78 I-95123 Catania, Italy}
\email{sergio.messina@oact.inaf.it; giuseppe.leto@oact.inaf.it; isabella.pagano@oact.inaf.it}


\begin{abstract}
Equal-mass stars in young open clusters and loose associations exhibit a wide spread of rotation periods, which likely  originates  from differences in the initial rotation periods and in the primordial disc lifetimes. 
We want to explore if the gravitational effects by nearby companions may play an additional role in producing the observed  rotation period spread, as well as, the role that magnetic activity may also play. We measure the photometric rotation periods of components of multiple stellar systems and look for correlations of the period differences among the components to their reciprocal distances. In this paper, we analysed the triple system AU Mic + AT Mic A\&B in the  25$\pm$3-Myr $\beta$ Pictoris Association. \rm  We have retrieved from the literature the rotation period of AU Mic (P =  4.85\,d) and measured from photometric archival data the rotation periods of   both components  of AT Mic (P  = 1.19\,d and P  = 0.78\,d) \rm for the first time. Moreover, we detected a high rate of flare events from AT Mic. Whereas the distant component AU Mic has evolved rotationally as a single star, the A and B components of AT Mic, separated by $\sim$27\,AU, exhibit a rotation rate a factor 5 larger than AU Mic. Moreover, the A and B components, despite have about equal mass,  show a significant difference ($\sim$40\%) between their rotation periods.  A possible explanation is that the gravitational forces between the A and B components of AT Mic (that are a factor $\sim$7.3$\times10^6$ more intense than those between AU Mic and AT Mic) have enhanced the dispersal of the AT Mic primordial disc, shortening its lifetime and the disc-locking phase duration, \rm making the component A and B of AT Mic  to rotate faster than the more distant  AU Mic. We suspect that a different level of magnetic activity between the A and B components of AT Mic may be the additional parameter responsible for the difference between their rotation periods.
\end{abstract}


\section{Introduction}
Low-mass  stellar members (M $<$ 1.2\,M$_\odot$)  of young open clusters and stellar association (age $<$ 0.5\,Gyr) exhibit a wide spread of their rotation periods. Within each cluster/association, we note that stars with similar masses have their rotation periods within a range of values. This range is minimum at early spectral types (from mid- to late-F)  and, generally, it increases when we move towards lower mass stars (see, e.g., Mamajek \& Hillenbrand 2008). This spread of rotation periods at the same stellar mass arises likely from differences in the initial rotation periods and in the primordial disc lifetimes. The shorter the disc lifetime, the shorter the disc-locking time, and the earlier the star begins to spinning up, owing to radius contraction (see, e.g., Camenzind 1990; Ribas et al. 2014).\\
We are carrying out a study to investigate  if gravitational effects can shorten the disc lifetime by comparing the rotation periods of close components of multiple systems with those of single stars. \rm
This study is part of the RACE-OC project (Rotation and ACtivity Evolution in Open Clusters; Messina 2007).
For this study, cluster or association stars that belong to triple or multiple systems are the best suited, especially if the components of these systems have similar masses. The origin of significantly different rotation periods among the components can reside in different initial rotation periods and/or in a different duration of the disc lifetimes, being similar all the other basic parameters (age, mass, metallicity). The architecture of the system, that is the reciprocal distance among the components, in these cases can play its own key role in differentiating the rotation periods. We have already analysed two such triple systems, BD$-$21\,1074 in the $\beta$ Pictoris Association (Messina et al. 2014), and TYC\,9300-0891-1AB/TYC\,9300-0525-1 in the Octans Association (Messina et al. 2016a). In both systems, there are two components on a wide orbit, and one having a nearby companion. In the first system (BD$-$21\,1074), we found that the nearby companion at 16\,AU significantly shortened the disc lifetime making one component to rotate significantly faster than the wide companion. In the second case (TYC\,9300-0891-1AB/TYC\,9300-0525-1), we found that  the nearby companion at 160\,AU was sufficiently distant  to have a negligible effect on the rotation, given that there was no period difference between the tight binary and the wide companion. \rm \\
Now, we present a third system,  AT Mic AB + AU Mic in the   25$\pm$3-Myr $\beta$ Pictoris Association (Messina et al. 2016b). AT Mic and AU Mic are at very large distance ($\sim$46200 AU) from each other, whereas AT Mic A and B are separated by only $\sim$27\,AU. We measured the photometric rotation periods of both components A and B. As we will show,  AU Mic has a rotation period comparable to those of other single stars and very wide components of binary systems, therefore we can assume that it has evolved as a single star with negligible external gravitational perturbation.  On the contrary, the close components of AT Mic with their rotation periods a factor 5 shorter than AU Mic, have likely reciprocally shortened their disc lifetime and  started to spin up earlier than the coeval AU Mic. \\ \rm
We also find that the difference between the rotation periods of the A and B components of AT Mic is significant, and explore the role that the flaring activity on one or both components may have played to produce such period difference.\\ 
In Sect.\,2, we present the literature information on AU Mic and AT Mic. In Sect.\,3 and 4, we present the photometric data and their periodogram analyses to measure the rotation periods of the AT Mic components. 
In Sect.\,5, we present our novel analysis on the flares detected on AT Mic. A discussion of the rotational properties of the components of the triple system is given in Sect.\,6. Conclusions are presented in Sect.\,7.

\section{Literature information}
AU Mic is a M1Ve single star (Matthews et al. 2015;  Bailey et al. 2012\rm).
It hosts a well-known debris disk that appears nearly edge-on and extends out to $\sim$200\,AU in radius (Kalas et al. 2004).
The rotation period P = 4.865\,d was discovered by Torres \& Ferraz-Mello \rm (1973). A new determination P = 4.854\,d was done by  Vogt et al. \rm (1983). A period P = 4.852\,d is reported by Kiraga (2012) and P = 4.822\,d is listed in All Sky Catalogue of Variable Stars (ACVS; Pojma\'nski 2002). 
A period P = 4.837\,d is found by Hebb et al. (2007).

AT Mic is a close visual binary consisting of two almost equal M4.5Ve + M4.5Ve components at a distance 
d = 10.70\,pc (Zuckerman \& Song 2004). Both stars show evidence for a significant
orbital motion. Wilson (1954) reports a separation $\rho$ = 3.6$^{\prime\prime}$ between the two components,  McCarthy \& White \rm (2012) report $\rho$ = 3.3$^{\prime\prime}$, whereas Riedel et al. (2014)  found the separation to vary  from $\rho$ = 2.8$^{\prime\prime}$ 
(PA = 171$^{\circ}$) in 2003 to  $\rho$ = 2.3$^{\prime\prime}$ (PA = 153$^{\circ}$) in 2012, the latter corresponding to a projected separation of $\sim$27\,AU,  and provided a more precise distance measurement d = 9.87$\pm$0.07\,pc. \rm
Preliminary orbital period P = 209\,yr and eccentricity  e = 0.26 are provided by Malkov et al. (2012).  \\
The membership of the $\beta$ Pictoris Association  was first proposed by Barrado y Navascu\'es et al. (1999), subsequently
confirmed by Zuckerman et al. (2001), Lepine \& Simon (2009), Malo et al. (2013, 2014), and Riedel et al. (2014).\\
Shaya \& Olling (2011) using Bayesian statistical methods found that AT Mic and AU Mic have very high (100\%) probability
to be a physical pair, despite their large angular separation ($\sim$4680$^{\prime\prime}$).
Therefore the close visual binary AT Mic together with AU Mic form a hierarchical triple system. 
The system has a reduced gravitational binding energy  U$^{\star}_g = -G\,M_{\rm A}\,M_{\rm B}\,s^{-1} = 8.9\times10^{33} J$ (Caballero 2009),   where M$_{\rm A}$ is the mass of AT Mic, M$_{\rm B}$ the mass of AU Mic, $G$ is the universal gravitation constant, and $s$ the projected separation. \rm This value lies at the boundary between very wide binaries and couples of single stars that are co-moving within the same stellar kinematic group (see, Alonso-Floriano et al. 2015; Caballero 2010).
However, as Caballero (2009) noted, this system is close to the bound limit and  close to be disrupted  by third bodies. \\
A number of projected rotational velocity measurements for AU Mic and for both the components of AT Mic are available in the literature and they are listed in Table~\ref{tab_vsini}.
\begin{table*}
\caption{\label{tab_vsini}AU Mic and AT Mic projected rotational velocity measurements from the literature. }
\begin{tabular}{clccl}
\hline
AU Mic & & AT Mic A & AT Mic B\\
\hline
$v \sin i$ & Ref. & \multicolumn{2}{c}{$v \sin i$}& Ref. \\

(km\,s$^{-1}$) & & \multicolumn{2}{c}{(km\,s$^{-1}$)}  & \\
\hline
8.2 & Weise et al. (2010)&  10.1 & 15.8 & Torres et al (2006)\\
$<$ 8.5 & Browning et al. (2010) & 10.56 & 17 & Scholz et al. (2007)\\
8.49 & Scholz et al. (2007) & 10.6 & 17 & Jayawardhana et al. (2006)\\
 9.3 & Torres et al. (2006) & 15$\pm$2 & 12$\pm$2 &  Lepine \& Simon (2009)\\
 &  & 13.6 & 8.7 & Reiners et al. (2012)\\
\hline
\end{tabular}
\end{table*}
We note that  Lepine \& Simon (2009) 
and Reiners et al. (2012),  quote the component A of AT Mic to rotate faster than the component B, contrary to the other authors. This later fact could be due to an exchange between  the A and B components.\\

As expected from the fast rotation and late spectral type, AT Mic is magnetically active. 
The system is listed in the catalogue of UV Cet-type flare stars (Gershberg et al. 1999). Flares were detected also in the X-ray wavelengths by XMM-Newton (Pye et al. 2015).\\
Plavchan et al. (2009) report on the detection of infrared excess at 24\,$\mu$m with the Spitzer telescope, whereas
Herschel observations did not detect any presence of either gas or dust disc (Riviere-Marichalar
 et al. 2014).  As well, Kennedy et al. (2014) did not find any flux excess at 24\,$\mu$m, explaining the difference with Plavchan et al. (2009) detection as due to the use of more recent model spectra,
where the photospheric flux is higher at the considered wavelengths.\\
Kiraga \& Stepien (2007) report on the measure of the photometric rotation period P = 0.78\,d of AT Mic from the analysis of the All Sky Automated Survey (ASAS; Pojma\'nski 1997) time series. Messina et al. (2011) report a different photometric period P = 1.197\,d from their analysis of SuperWASP photometry (Butters et al. 2010). Neither ASAS nor SuperWASP photometry could spatially resolve the two components, providing only integrated fluxes.\\
Since the rotational properties of AU Mic have been well established in the mentioned earlier studies, in the present investigation, we carry out a deeper analysis of the SuperWASP photometry, which also includes a flare analysis, with the aim to better characterize the rotational properties of the two components A and B of AT Mic.


\section{Data preparation}
AT Mic is a SuperWASP target with identification number 1SWASPJ204151.14-322610.2. From the public archive 
we could retrieve observations collected in 2006, 2007, 
and in a few days of 2008. A visual inspection of the data timeseries, subsequently confirmed by a statistical analysis,
clearly showed that the quality
of the observations starting from late 2006 degraded significantly, owing to the presence of systematic errors that
remained uncorrected despite the application of the SysRem algorithm of Tamuz et al. (2005).

\begin{figure}
\begin{minipage}{10cm}
\centerline{
\includegraphics[width=70mm,height=90mm,angle=90,trim= 0 0 0 100]{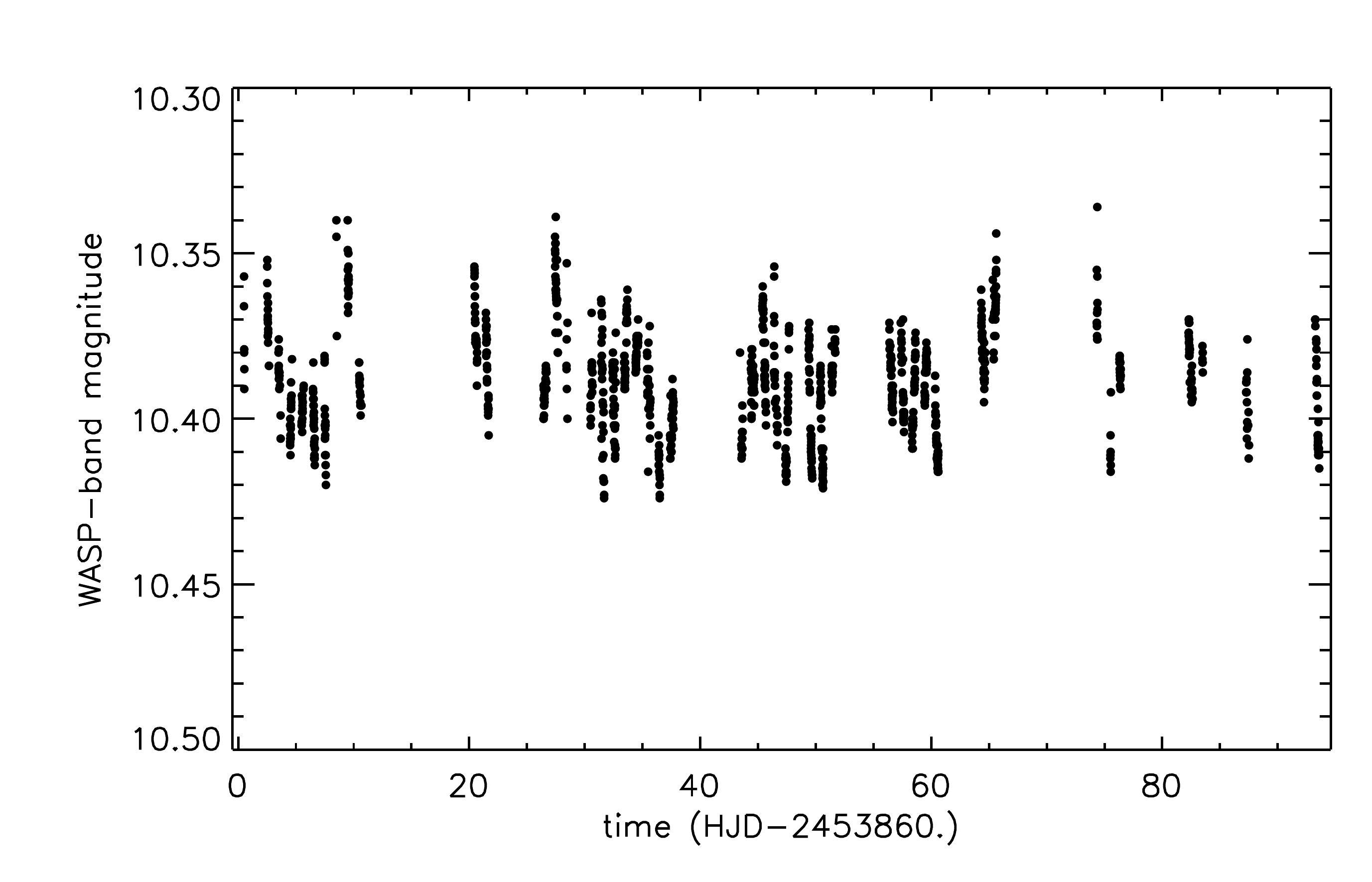} \\
}
\end{minipage}
\caption{\label{timeseries}SuperWASP magnitude time series of AT Mic. Flare events were removed for the rotation period search.}
\vspace{0cm}
\end{figure}

For this reason, we focussed our analysis on the higher-quality data collected during  2006. More specifically,
we selected data from May, 5 to August 6, 2006 for a total of 3097 observations.  The high cadence of the SuperWASP observations allowed us to identify with visual inspection numerous flare events thanks to their typical shape consisting of a rapid brightness rise followed by an exponential-like fading to the quiet state. All these events were removed from the time series. Then, we applied a moving boxcar filter with 
3$\sigma$ threshold to remove any residual outliers. \rm
 Finally, we averaged consecutive magnitudes using a bin width of 8 minutes (corresponding
on average to two-three consecutive measurements). In this way, we were left with 1062 average data points for the subsequent periodogram analysis 
(see Fig.\ref{timeseries}), and 300 points,  which were left un-averaged to preserve the highest temporal resolution, collected during flaring events for the subsequent
flare analysis.

\begin{figure*}
\begin{minipage}{20cm}
\centerline{
\includegraphics[width=90mm,height=120mm,angle=90,trim= 0 0 0 0]{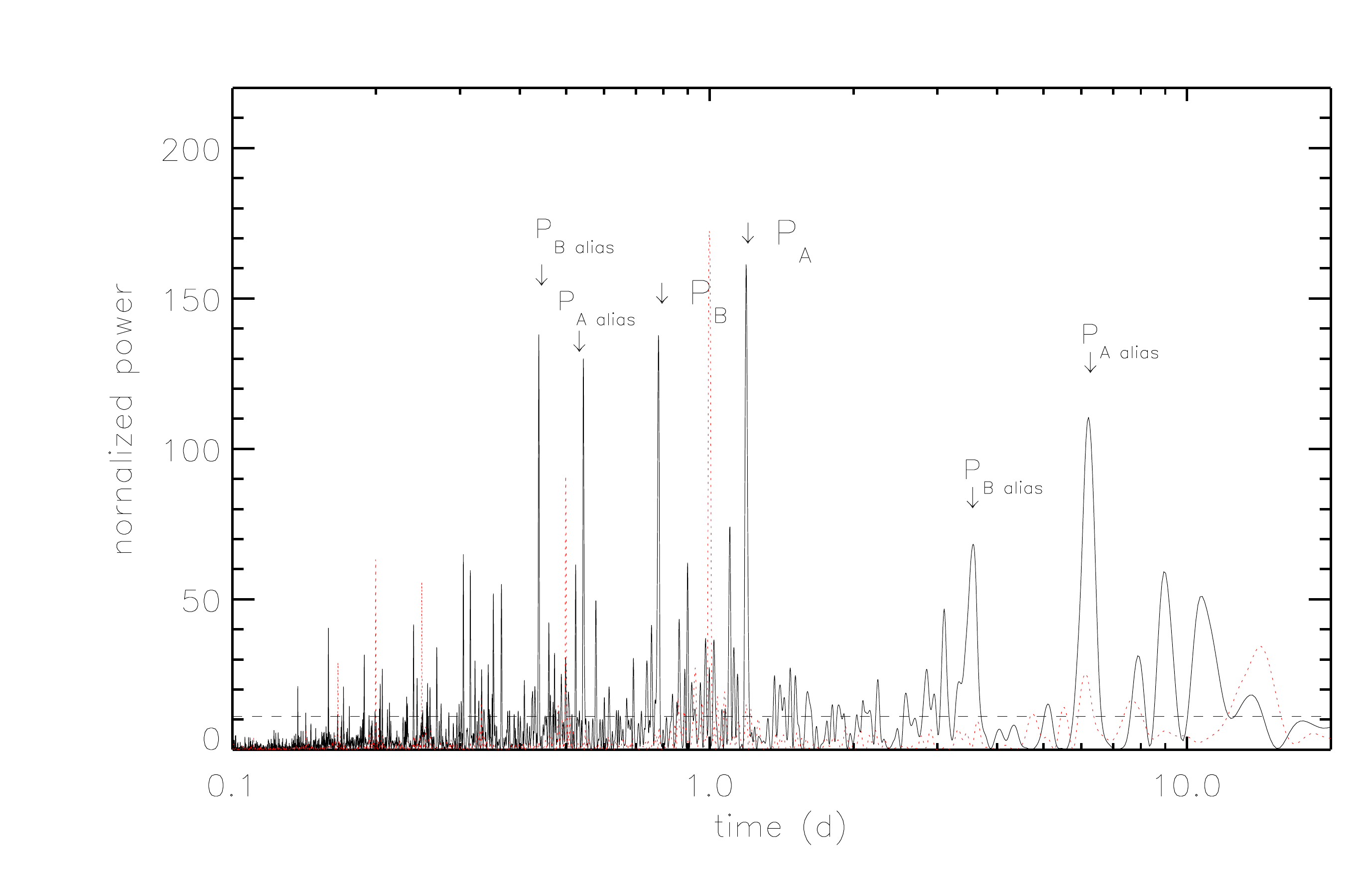} \\
}
\end{minipage}
\caption{\label{periodogram_ls}LS periodogram of AT Mic time series. We mark the rotation periods P=1.19\,d and P = 0.78\,d of the components of AT Mic, and their alias periods. The dashed line indicates the power level corresponding to a confidence level of 99\%, whereas the dotted red line indicates the spectral window. The subscript indices A and B indicate the primary and secondary periods detected by the periodogram analysis.}
\vspace{0cm}
\end{figure*}  

\begin{figure*}
\begin{minipage}{20cm}
\centerline{
\includegraphics[width=90mm,height=120mm,angle=90,trim= 0 0 0 0]{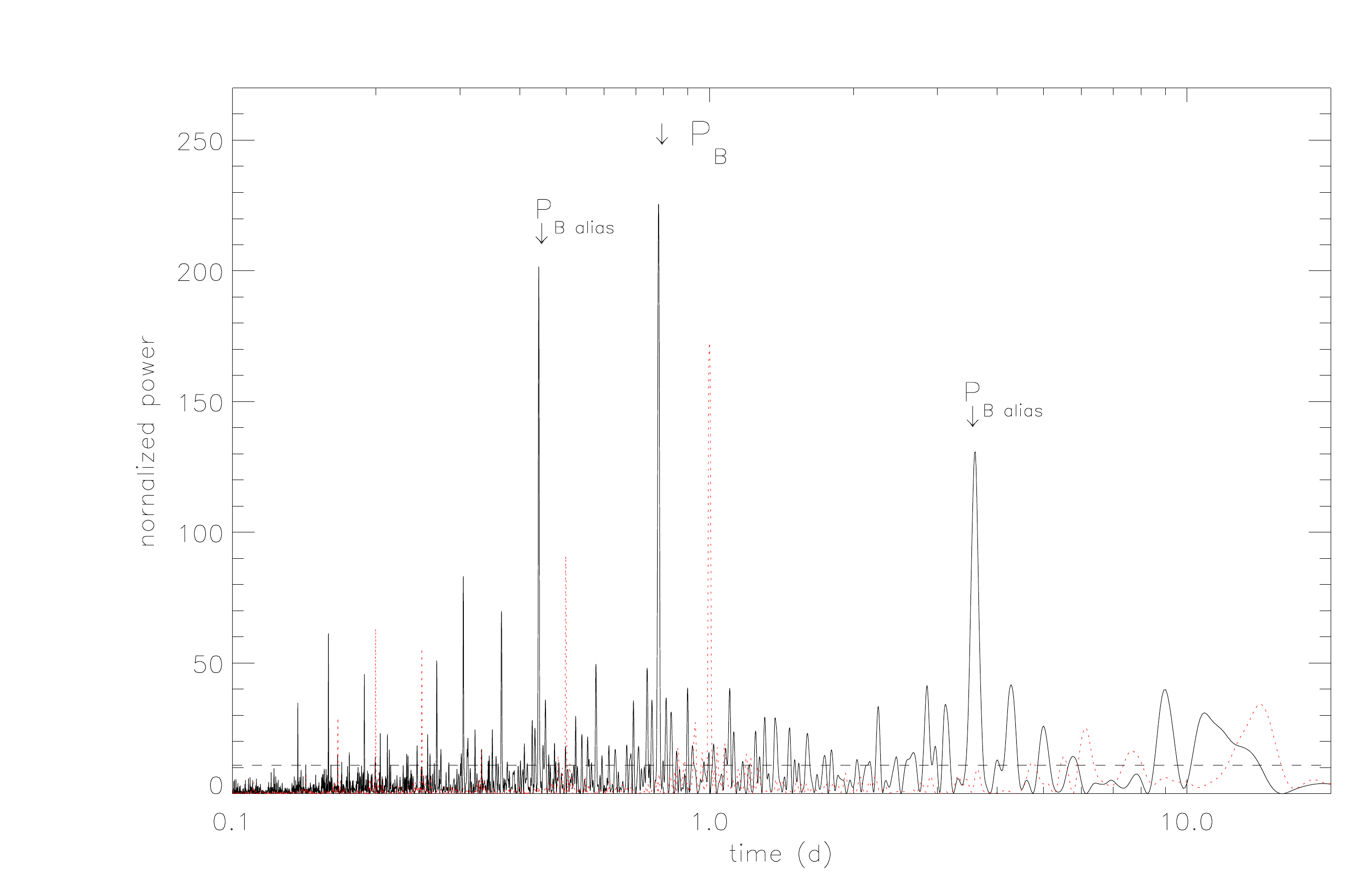} \\
}
\end{minipage}
\caption{\label{periodogram_ls_filt}Same as Fig.\,\ref{periodogram_ls}, but with the primary frequency (corresponding to P = 1.19\,d) having been 
filtered out.}
\vspace{0cm}
\end{figure*}

\section{Periodogram analysis}
\begin{figure*}
\begin{minipage}{20cm}
\centerline{
\includegraphics[width=90mm,height=120mm,angle=90,trim= 0 0 0 0]{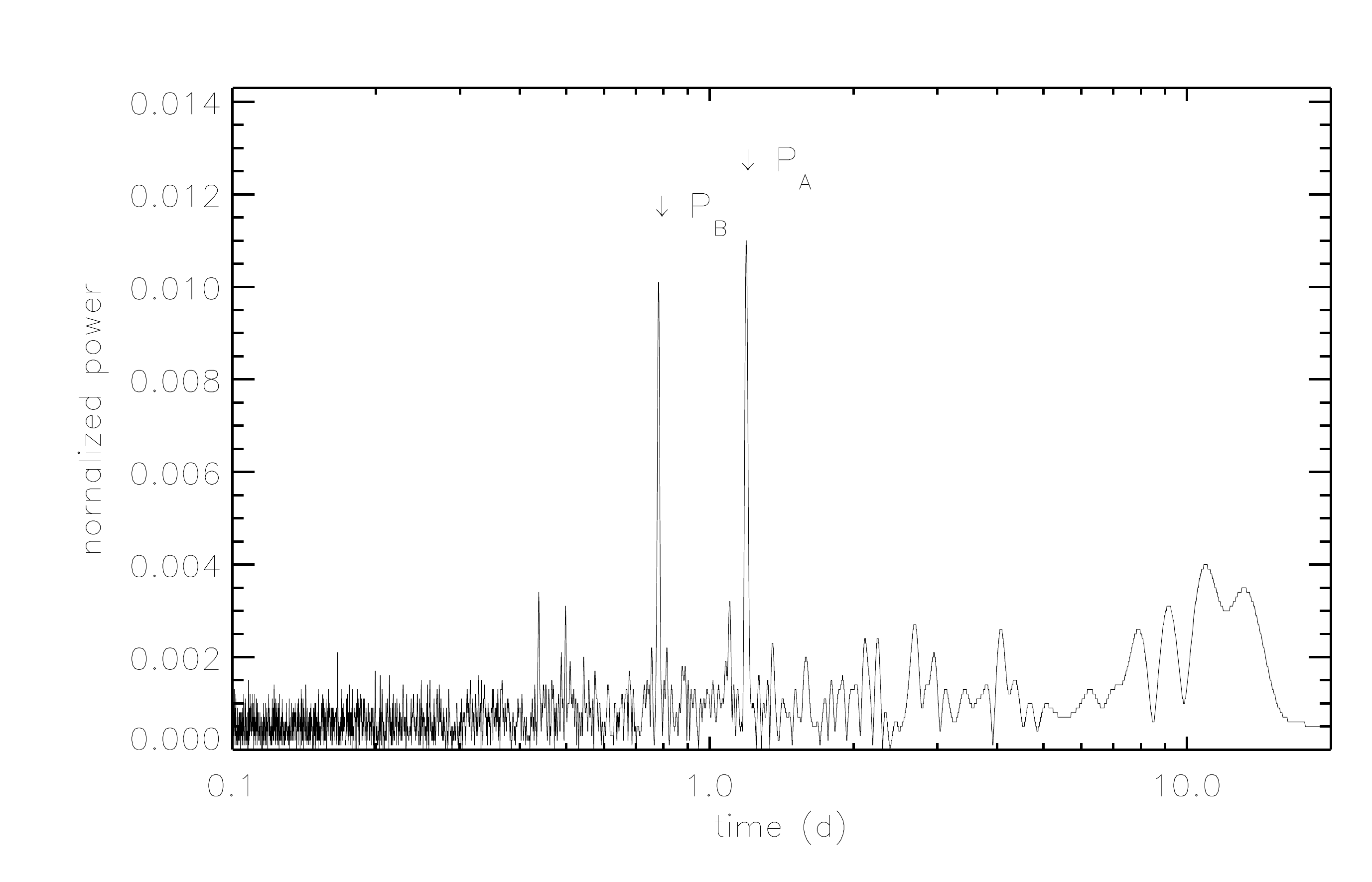} \\
}
\end{minipage}
\caption{\label{periodogram_clean} Results from Clean periodogram analysis, where the alias periods have been effectively removed.}
\vspace{0cm}
\end{figure*}

The two M4.5 components of the AT Mic system have a separation of about 3$^{\prime\prime}$. Therefore, all SuperWASP observations provide only the integrated magnitudes (the minimum aperture radius for the magnitude extraction being $\sim$ 35$^{\prime\prime}$). We used the Lomb-Scargle (LS; Scargle 1982)
and Clean (Roberts et al. 1989) algorithms to search for the rotation periods of both components. In fact, the system's components have similar brightness and, having same spectral type and similar projected rotational velocities, are expected to 
exhibit similar activity levels. Therefore, we expect that they contribute almost equally to the observed variability.
In Fig.\ref{periodogram_ls}, we plot the results of the Lomb-Scargle periodogram analysis. The solid line is the periodogram whereas the dotted red line is the window
function related to the data sampling. The horizontal dashed line indicates the power level corresponding to a False-Alarm-Probability FAP =  0.1\%,  that is the probability that a power peak of that height simply arises from Gaussian
noise in the data. The FAP was estimated using a Monte-Carlo method, i.e., by generating
1000 artificial light curves obtained from the real one, keeping the date but scrambling the
magnitude values (see, e.g., Herbst et al. 2002). The uncertainties of the rotation periods is
computed following the prescription of Lamm et al. (2004).\\
We identified six major power peaks at very high significance level. The dominant peak is at a period P$_A$ = 1.19\,d. This period allows us to compute its possible
beat periods, which  arises from the 1-d data sampling imposed by the fixed longitude of the observation site and the day-night duty cycle imposed by the Earth's rotation.
Such beat periods are P$_{\rm A\_alias 1}$ = 6.26\,d and P$_{\rm A\_alias 2}$ = 0.54\,d. The second major peak is at P$_B$ = 0.78\,d and its beat periods are 
P$_{\rm B\_alias 1}$ = 3.54\,d and P$_{\rm B\_alias 2}$ = 0.44\,d.\\
If P$_A$ = 1.19\,d is the correct rotation period of one of the two components, we expect that filtering it out from the time series and recomputing the periodogram,
it and its beat periods must disappear. Indeed, that is what we found as shown in Fig.\,\ref{periodogram_ls_filt}. The remaining power peaks are those corresponding to 
P$_B$ and its beats.\\
\indent
The Clean algorithm, differently than Lomb-Scargle, effectively removes the effect of the spectral window from the resulting periodogram. In Fig.\,\ref{periodogram_clean} 
only the rotation periods corresponding to the two components are left, whereas any beat period is effectively removed.

 The significant difference ($\sim$40\%) between these two periods rules out the possibility that we may have detected 
the signal from two major spot groups at different stellar latitudes on a differentially rotating star. Moreover, as discussed in the next section, these two values agree well with the projected rotational velocities of the two components, when the spin/orbit alignment is assumed. Therefore, we are confident that the periodicities we found from  our periodogram analysis are the rotation periods of the two components:  P$_A$ = 1.19$\pm$0.03d and P$_B$ = 0.78$\pm$0.01d,   where the subscript indices A and B indicate the primary and secondary periods detected by the periodogram analysis. \rm
\rm 

\begin{figure}
\begin{minipage}{10cm}
\centerline{
\includegraphics[width=70mm,height=90mm,angle=90,trim= 0 0 0 80]{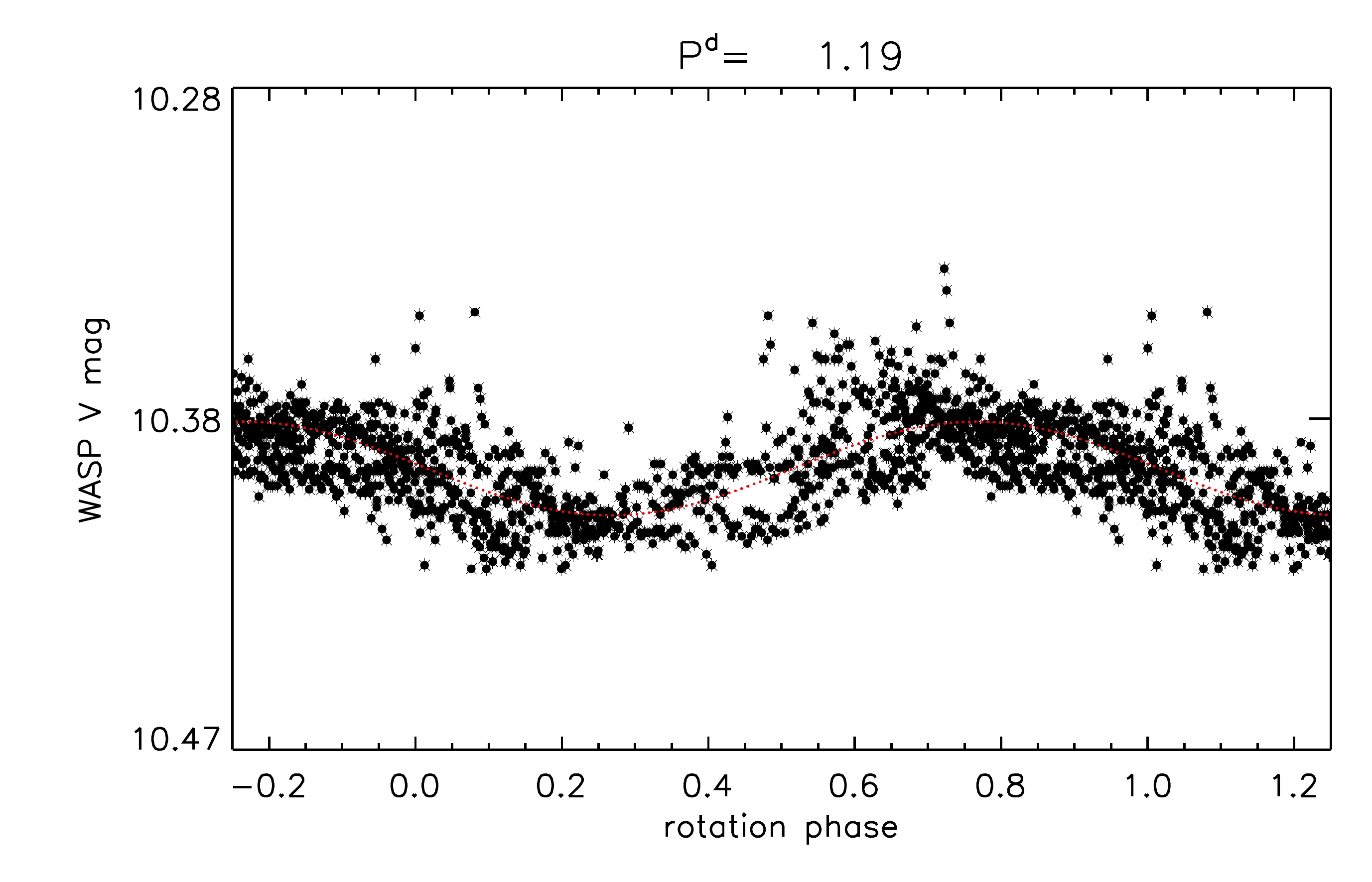} \\
}
\end{minipage}
\caption{\label{period1}Phased light curve of AT Mic with the period P = 1.19\,d, where the periodicity P = 0.78\,d has been filtered out.}
\vspace{0cm}
\end{figure}

\begin{figure}
\begin{minipage}{10cm}
\centerline{
\includegraphics[width=70mm,height=90mm,angle=90,trim= 0 0 0 80]{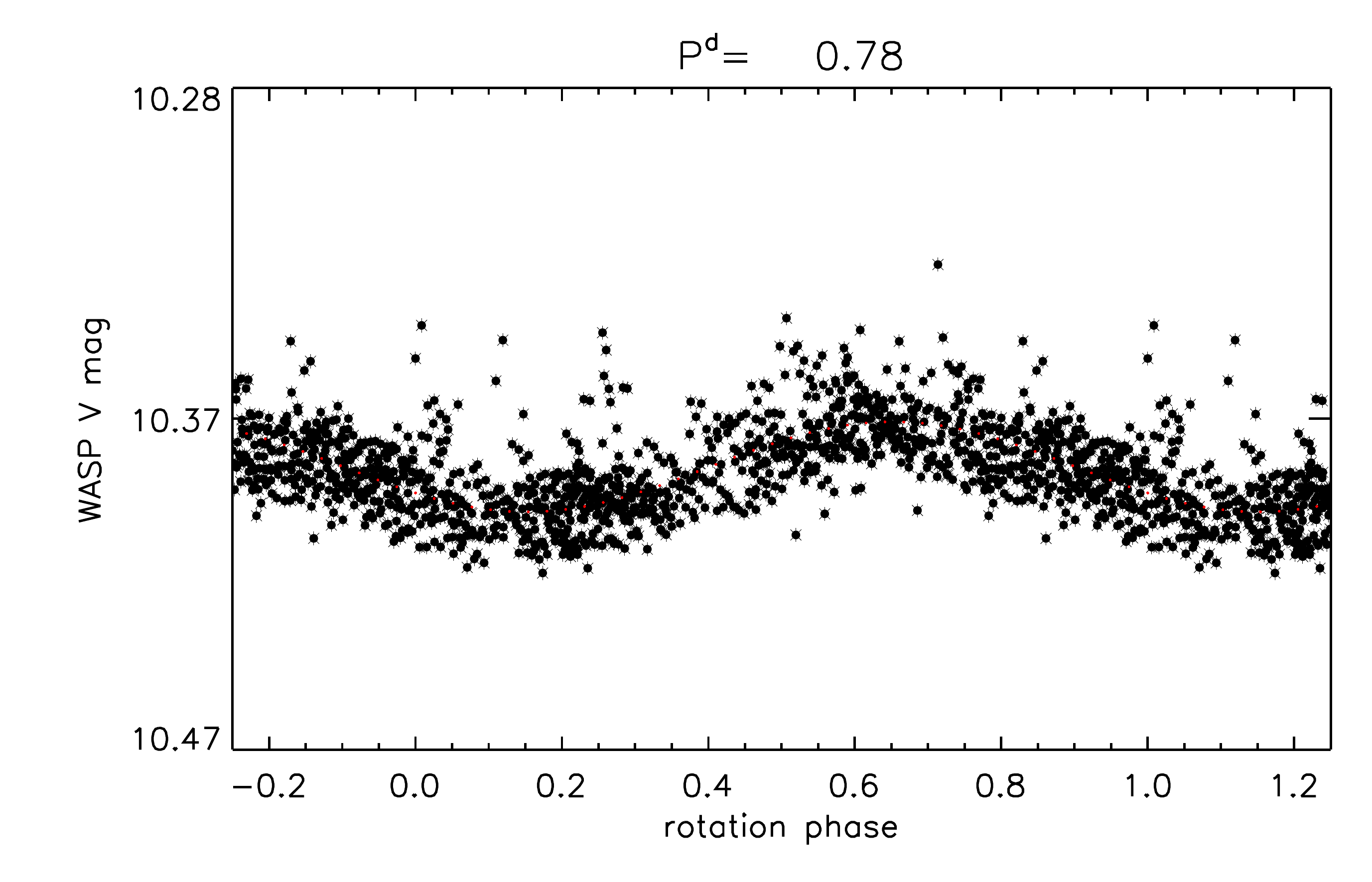} \\
}
\end{minipage}
\caption{\label{period2}Phased light curve of AT Mic with the period P = 0.78\,d, where the periodicity P = 1.19\,d has been filtered out.}
\vspace{0cm}
\end{figure}

In Fig.\,\ref{period1}, we plot the phased light curve with the rotation period P = 1.19\,d (after filtering out the  secondary periodicity P = 0.78\,d from the time series). In Fig.\,\ref{period2}, after filtering  the primary periodicity P = 1.19\,d, we
plot the phased light curve with the period P = 0.78\,d. 

\section{The AT Mic A \& B parameters}
Using the distance d = 9.87\,pc and deblended V magnitudes from Riedel et al. (2014), bolometric corrections and effective temperatures
from Pecaut \& Mamajek (2013), corresponding to the M4.5 spectral type, we derive the bolometric magnitudes, luminosities
and radii for  AT Mic A and B (see Table\,\ref{parameters}).

\begin{table}
\caption{\label{parameters}Stellar parameters of AT Mic A \& B. See text for the discussion on the determination of the spin axis inclination.}
\begin{tabular}{ccc}
\hline
 & AT Mic A & AT Mic B\\
 \hline
d (pc) & 9.87$\pm$0.07 & 9.87$\pm$0.07\\
V (mag)  & 11.09$\pm$0.02  & 11.13$\pm$0.02\\
BC$_{\rm V}$ &  -2.82 & -2.82\\
M$_{\rm bol}$ & 8.29$\pm$0.05 & 8.33$\pm$0.05\\
L  (L$_\odot$) & 0.037$\pm$0.004 & 0.035$\pm$0.004\\
R (R$_\odot$) & 0.72$\pm$0.08 & 0.71$\pm$0.08\\
 \multicolumn{3}{c}{Case A}\\
 P (d)   & 1.19	& 0.78\\
 $<v\sin{i}>$ (km\,s$^{-1}$)  & 10.4 & 15.7\\
i ($^{\circ}$)     &20$\pm$3 & 20$\pm$3 \\
\multicolumn{3}{c}{Case B}\\
P (d) & 0.78 & 1.19 \\
 $<v\sin{i}>$ (km\,s$^{-1}$)  & 15.7 & 10.4\\
i ($^{\circ}$)     & 20$\pm$3 & 20$\pm$3 \\
\hline
\end{tabular}
\end{table}

Combining rotation period, radius, and projected rotational velocity we can derive the inclination of the stellar rotation axis for both components. However, we do not know a priori to which component the measured rotation periods refer. To restrict the possible combinations, we make the assumption that both components have same inclination of their spin axis. This assumption arises from the consideration that the two components are physically bound and relatively close to each other at a distance of $\sim$27\,AU and sufficiently young to consider negligible any effect on the inclination from stellar encounters. Therefore, it is reasonable to expect that both stars have their equatorial planes aligned with their common orbital plane. 
Since R$_{\rm A}$/R$_{\rm B}$ $\simeq$ 1 and, assuming  $\sin{i}_A$/$\sin{i}_B$ $\simeq$ 1, then P$_{\rm A}$/P$_{\rm B}$ = $v\sin{i}_B$/$v\sin{i}_A$. This allows only two possible cases for the A and B components: case A) we assign to the A component  P$_{\rm A}$ = 1.19\,d and $<v\sin{i}_A>$ = 10.4\,km\,s$^{-1}$, whereas  to the B component P$_{\rm B}$ = 0.78\,d and $<v\sin{i}_B>$ = 15.7\,km\,s$^{-1}$; and case B )  we assign to the A component P$_{\rm A}$ = 0.78\,d and $<v\sin{i}_A>$ = 15.7\,km\,s$^{-1}$, and to the B component P$_{\rm B}$ = 1.19\,d and $<v\sin{i}_B>$ = 10.4\,km\,s$^{-1}$.
In both cases we derive  $i_{\rm A}$ = $i_{\rm B}$ = 20$\pm$3$^\circ$ and conclude that the projected rotational velocity measurements by Lepine \& Simon (2009) and Reiners et al. (2012) were effectively exchanged between the two components.
Unfortunately, we have no additional information to assign the correct rotation periods to the AT Mic components, since Case A and B are equally possible.\\
We know that the amplitude of the light curve depends on a combination of
the fraction of photosphere covered by starspots asymmetrically distributed in longitude, their latitude, 
and the inclination of the stellar rotation axis. For a given amount of spots, the smaller the value of $i$,
the smaller the light curve amplitude. In our case, despite the high activity level of the two M4.5V components,
as manifested by the high frequency of flare events, and as expected from the fast rotation rate and late spectral type, we
measure a light curve amplitude not larger than 0.03\,mag in the SuperWASP V band. This value can increase up to 0.06\,mag if we take into account the dilution effect arising from the integrated photometry. Even this corrected value is quite small for a very active star, but very well consistent with the low inclination  $i$ = 20$^\circ$  of the rotation axis of both stars. \rm
Finally, we note that the level of magnetic activity usually changes in time owing to activity cycles and/or long-term trends.
Therefore, there can be epochs when the activity of one component is dominant with respect to the activity of the other component. This may explain why Kiraga (2012) analyzing the ASAS photometry detected only the rotation period of one component, while in the 2006 SuperWASP photometry, 
we were able to detect the rotational induced variability of both components, being the activity levels of the AT Mic components comparable (peak-to-peak amplitudes of the light curves are $\Delta$V = 0.027\,mag and $\Delta$V = 0.025\,mag).

 \section{Flare data reduction and analysis} 
\label{flaredata}

   Our data time series extends from HJD = 2453860.5009 to  HJD = 2453953.5930. Although data are not regularly spaced  nor complete, along the time series  we were able to detect  11 \rm clear flare episodes.  The time series spans 93.0921 days  (2234.21 hours) with a total time coverage of 10.7690 days (258.45 hours). The total time covered by detected flare episodes is  2.938 days (70.52 hours), \rm i.e.,  the system AT Mic was detected in a  flaring state  for  about 25${\%}$ of the observed time. Data were reduced following the procedure  described in Leto et al. (1997).\\ 
 The measured flare parameters are given in Table \ref{flaresdata}.  The time of the observed maximum is a good approximation of the  time at which the flare peak emission was reached for only 5 over  11 events.  \rm In fact,  6 flares \rm
 have their maximum  outside the available time series. In such a case, the measured flare amplitude, peak  luminosity and emitted energy are lower limits.
   The flare luminosity  is derived assuming  a quiescent emission level for AT Mic  of  L = 2,746e+30\,erg\,s$^{-1}$, \rm  as derived from flux calibration taking into account a distance d = 9.87$\pm$0.07\,pc, and a value V = 10.25\,mag for the quiescent magnitude of AT Mic (Koen et al. 2010). 

The equivalent duration (ED) given in Table~\ref{flaresdata} is the time needed to the quiescent AT Mic to produce the same amount of energy released in excess during the flare, and it is evaluated as follows:
\begin{equation}
ED = \sum_{}^{}   {\frac{(I-I_0)}{I_0} \times {\Delta t} }
\end{equation}
where $I$ is the light curve intensity and $I_0$ the intensity level out of flare, respectively. 
The energy is then estimated as E = L $\times$ ED (Leto et al. (1997).
\begin{table*}
\caption{Flare measured parameters: HJD is the Heliocentric Julian Date  of the observed maximum for each event. Up arrows annotate flares having their maximum outside the available time series. For these events italic font is used for the flare magnitude, energy, maximum luminosity  and equivalent duration that are lower limits for the true values.}
\begin{center}

\begin{tabular}{|r|c|r|c|c|}  
\hline
  &   &  & Max & Equivalent     \\
 HJD & $\Delta$mag   & Energy &  luminosity & duration \\
  & &  10$^{32}$ $erg$ &10$^{30}$ $erg\,s^{-1}$ & $min$ \\
\hline
 2453864.0318 & 0.044 $\pm$ 0.002 & 2.131 & 2.8 &    1.29 \\ 
 2453897.6362 & 0.029 $\pm$ 0.005 & 1.226 & 2.8 &    0.74  \\ 
$\uparrow$ 2453903.4896 & {\it 0.056} $\pm$ 0.017 & {\it 8.073} & {\it 2.9} &    {\it 4.90}  \\ 
$\uparrow$  2453911.5554 & {\it 0.038} $\pm$ 0.003 &{\it  6.068}&{\it 2.8 }&    {\it 3.68}  \\
2453916.7165 & 0.026 $\pm$ 0.003 & 0.868 & 2.83 &    0.53  \\ 
$\uparrow$ 2453918.0567 & {\it  0.025} $\pm$ 0.003 & {\it 0.493}& {\it 2.8 }&   {\it  0.30}  \\ 
$\uparrow$  2453923.8298 & {\it 0.032} $\pm$ 0.024 & {\it 5.512} & {\it 2.8} &   {\it  3.34} \\ 
$\uparrow$ 2453936.1268 & {\it 0.106} $\pm$ 0.002 & {\it 15.090} &{\it  3.0} & {\it    9.16}  \\ 
$\uparrow$ 2453943.3022 & {\it 0.062} $\pm$ 0.004 & {\it 9.730}  &{\it  2.9} &  {\it   5.90}  \\ 
$\uparrow$ 2453944.0122 & {\it 0.031} $\pm$ 0.002 & {\it 2.247}  & {\it 2.8} &    {\it 1.36} \\ 
2453953.2609 & 0.045 $\pm$ 0.011 & 8.579 & 2.8&    5.21  \\ 
\hline
\end{tabular}
\end{center}
\label{flaresdata}
\end{table*}%

\begin{figure*}
\centering
\includegraphics[width=1.1\textwidth,clip,origin=l]{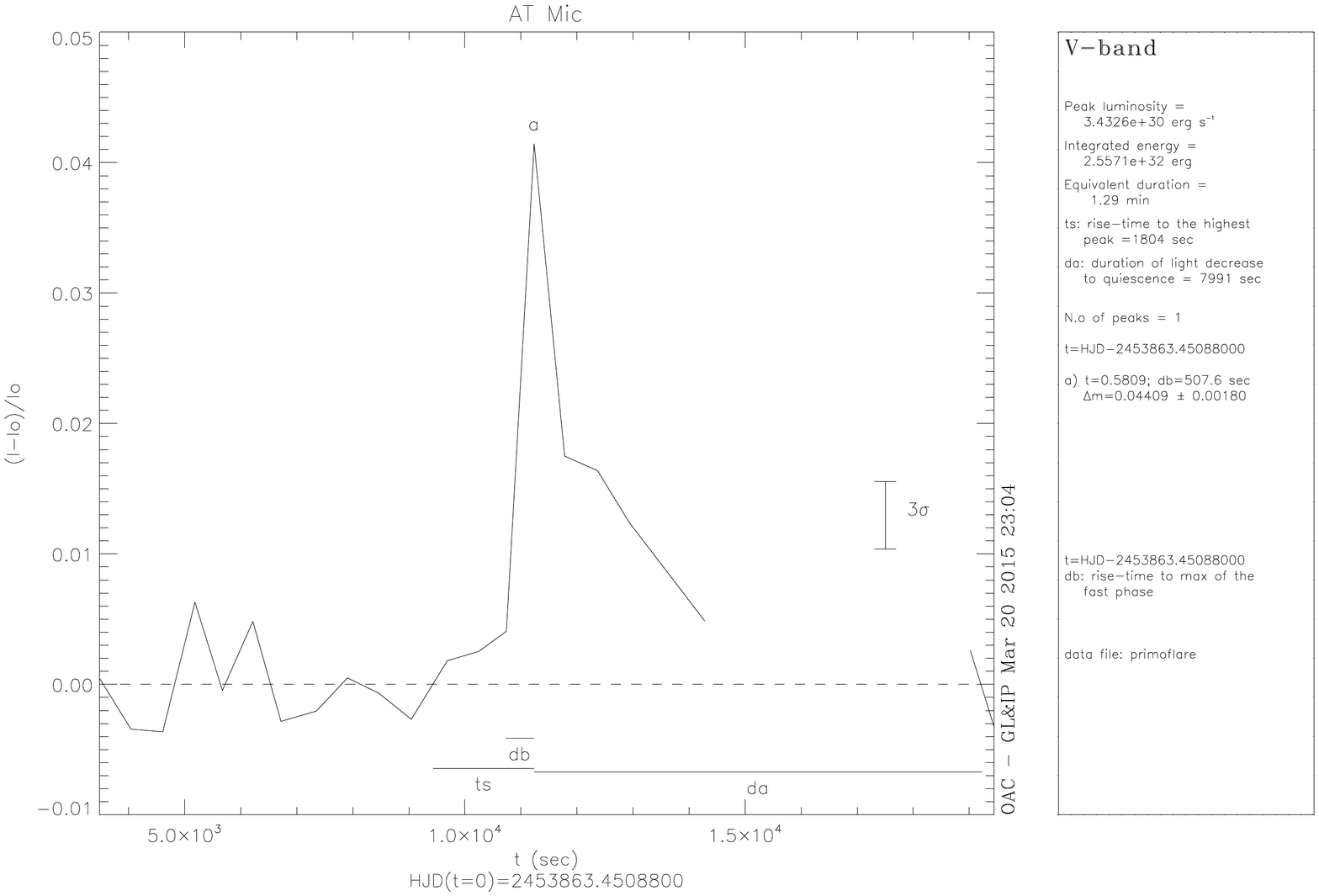}
\caption{Example of flare observed at HJD = 2453864.0318.}
\label{primoflare}       
\end{figure*}

Figure~\ref{primoflare} shows, as an example,  one of the observed  flares.

\begin{figure*}
\centering
\includegraphics[width=90mm,height=85mm,angle=0,trim= 0 0 0 0]{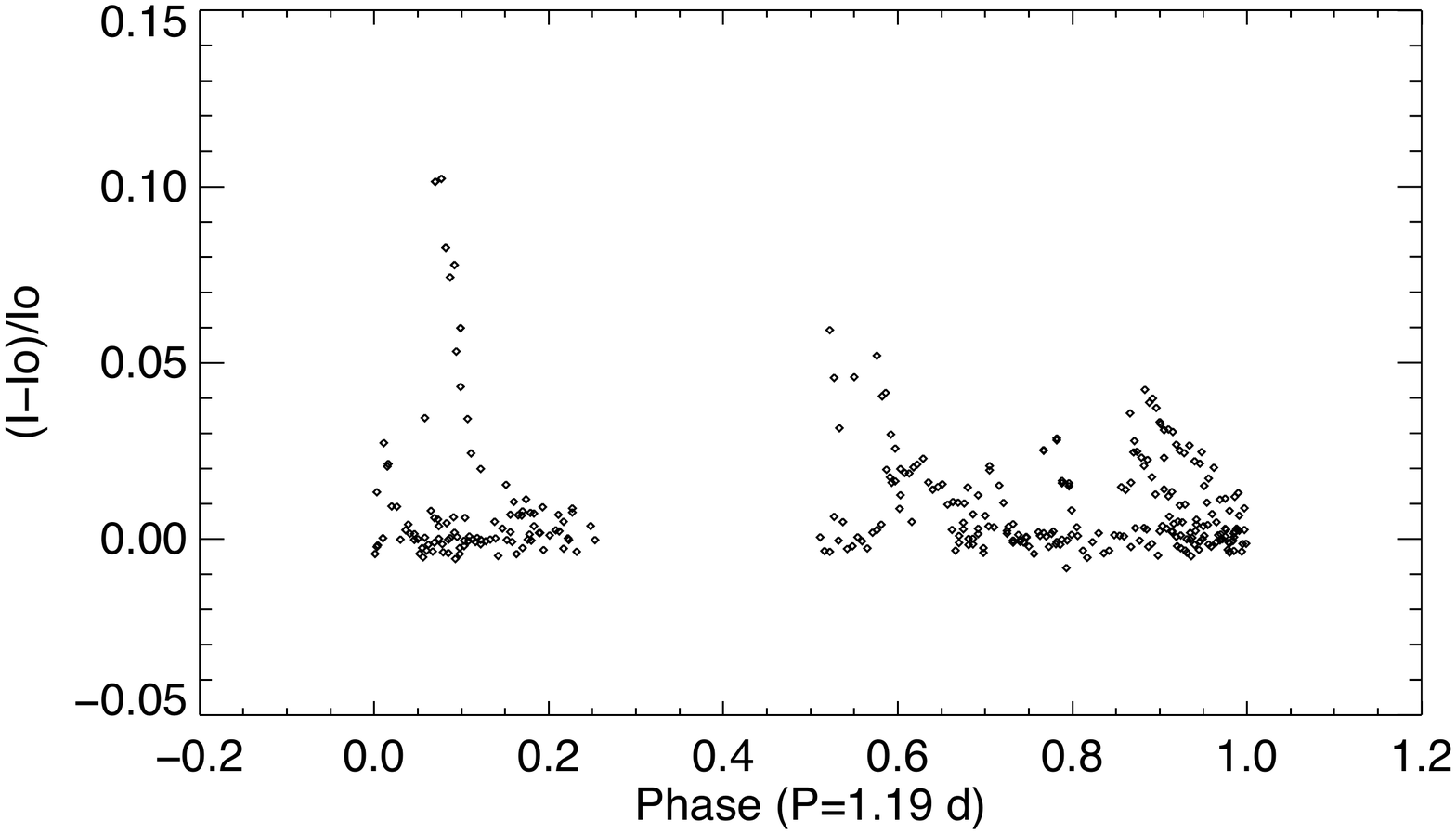}
\includegraphics[width=90mm,height=85mm,angle=0,trim= 0 0 0 0]{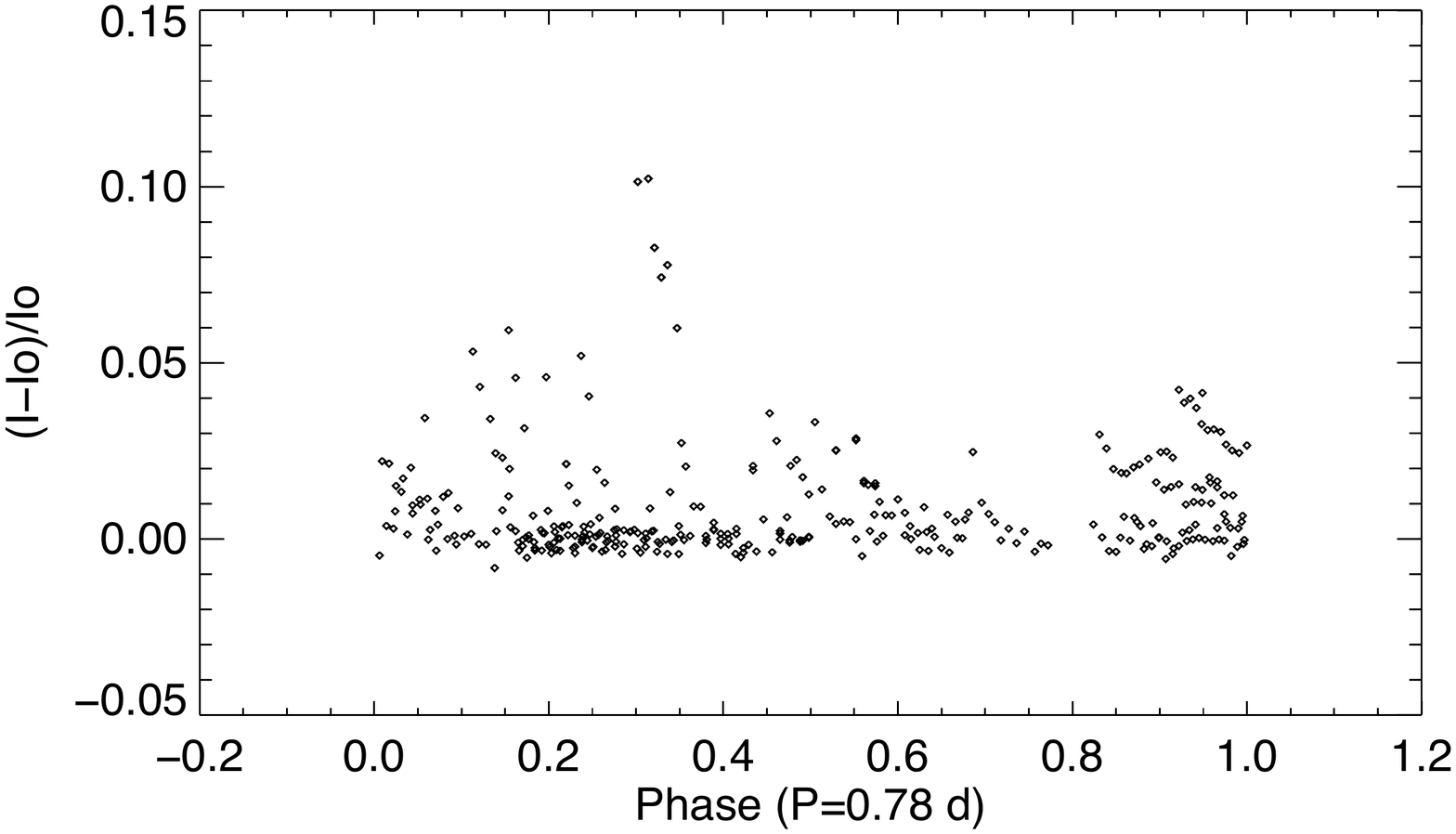}
\caption{Data acquired during flares (normalized intensity) folded in phase with the period P$_A$ = 1.191d (left panel) and with the period P$_B$ = 0.78d (right panel).}
\label{flarep1}       
\end{figure*}

As mentioned, the analysed photometric time series shows the occurrence of 12 flares, some of them complete, other only partially recorded. 
 To search for any preferred phase for flare occurrence along the star rotation, the  subset of data acquired during flares have been folded in phase by using the two  periods derived by the analysis of quiet-state data (see Sect.\,4).  Data folded with P$_{\rm A}$ = 1.19\,d (see left panel of Fig.\,\ref{flarep1}) show a  gap of flaring activity  between phases 0.25 and 0.5.  Instead, flaring data phased by using the period P$_{\rm B}$ = 0.78\,d do not show any particular concentration or gaps (see right panel of Fig.\,\ref{flarep1}).

\begin{figure}
\centering
\includegraphics[width=.5\textwidth,clip,origin=l]{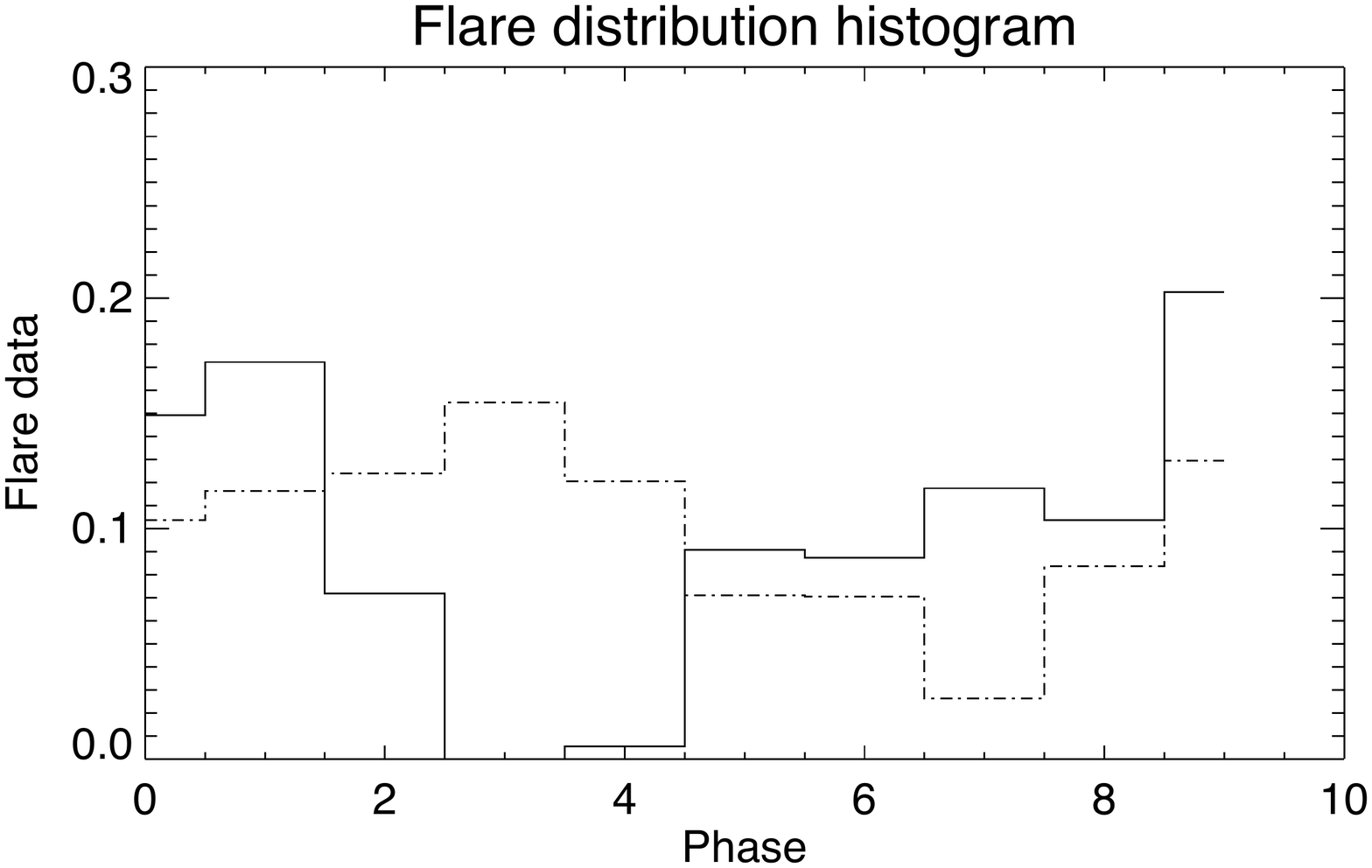}
\caption{Normalized distributions of flare events vs. rotation phase for P = 1.19\,d (solid line) and  P = 0.78\,d (dashed line).}
\label{histoflares}       
\end{figure}

To check for a possible bias due to a not random distribution of the observed epochs in the full time series, we have binned the full data set and the selected flare data set,  phase bins was 0.1. The resulting histograms for flare data were normalised to the histogram obtained for the full data sets in order to remove any possible bias, and to the summation of all 10 densities per bin to normalise to 1 the integral. The histograms are showed in Figure \ref{histoflares},  where the solid line represents the flare data set folded by using period P$_A$ as described and dashed line the same set similarly folded by using P$_B$. Flares show a correlation with the light curve phased with the longer period (the minimum flare frequency corresponds to the light curve minimum, when spots are best in view), the histogram folded with the shorter period  does not show evidence of correlation instead.
However, because of the small number of flare events considered in this analysis, the differences from bin to bin are not highly significant. Therefore, we are not in the position to establish if both components or only one are responsible for the observed flares.

 \section{Rotation period distribution}
\begin{figure*}
\centering
\includegraphics[width=0.6\textwidth,clip,origin=l,angle=90]{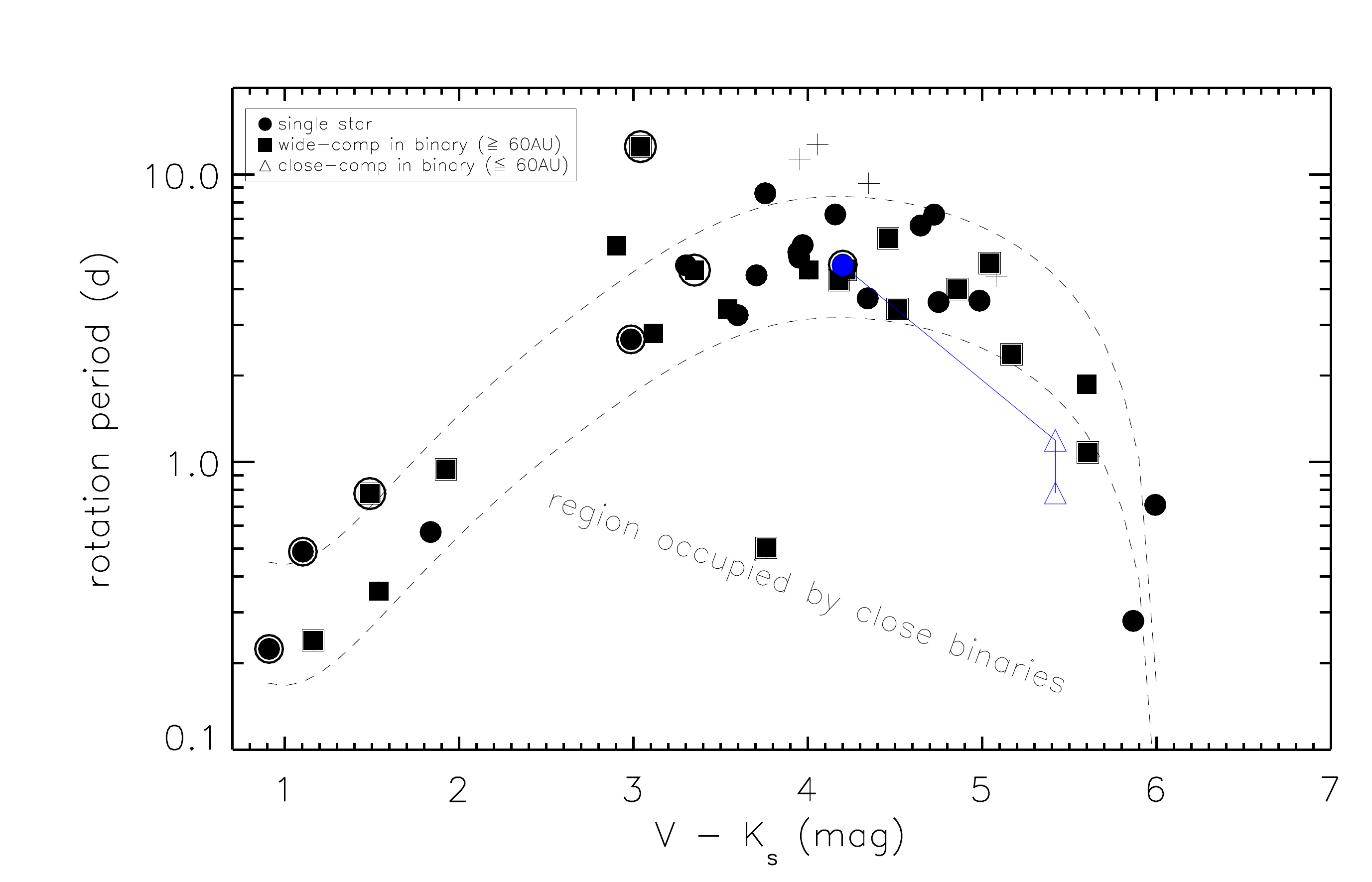}
\vspace{-7cm}
\caption{Distribution versus V$-$K$_s$ color of the rotation periods of the $\beta$ Pictoris  bona fide \rm  single members (filled bullets) and components of wide ($>$60\,AU) binary systems (filled squares) (Messina et al. in preparation).  Circled symbols are stars hosting debris discs. Dashed lines represent polynomial fits to the upper and lower bounds of the period distribution. AU Mic, and  AT Mic A \& B are represented with a blue circled bullet and open triangles, respectively, connected by solid lines.}

\label{distri}       
\end{figure*}

 The AT Mic AB + AU Mic triple system is key to investigate the origin  of rotation period spread generally observed among coeval stars at young ages.\\
Color-period diagrams of young clusters/associations display an upper bound in the distribution whose bluer-part (consisting of stars,   starting from the F spectral type, \rm  that first settle on the ZAMS)
moves   progressively \rm toward longer rotation periods and with a decreasing 
dispersion as far as the stellar age increases. 
By an age of about 0.6\,Gyr, F-G-K stars exhibit an almost one-to-one correspondence 
between rotation period and mass, as in the case of Hyades (Delorme et al. 2011), Praesepe (Douglas et al. 2014) and Coma Berenicis (Collier Cameron et al. 2009)
open clusters. Such an univocal dependence is currently exploited for gyro-chronological estimate of stellar age (see, e.g., 
Mamajek \& Hillenbrandt 2008).\\
Within the same stellar cluster, the distribution of rotation periods depends on mass and, for any mass bin, it depends on the 
initial rotation period and on the angular momentum evolution that can vary from star to star. \\
In our specific case, AU Mic has a rotation period P = 4.83\,d. In contrast, the components of AT Mic exhibit a much faster rotation with 
 P$_{\rm A}$ = 1.19\,d and P$_{\rm B}$ = 0.78\,d. Thus,  AU Mic has a rotation period a factor 5 longer than the average period of the AT Mic components. Moreover, the components of AT Mic, although equal in mass, age, and metallicity, have a significant difference between their rotation periods of about 40\%.\\
 	In Fig.\,\ref{distri}, a comparison with the period distribution of bona fide members of the $\beta$ Pictoris association shows that the difference between the rotational properties of AU Mic and AT Mic is partly due to their different masses (M1V against M4.5V, respectively), since the rotation period distribution is mass dependent.
 However, we note that AT Mic A and B rotate still faster than equal-mass/color single stars or wide components of multiple systems. In fact, their rotation periods are outside the 3$\sigma$ distribution marked by the dashed lines in Fig.\,\ref{distri} and fall in the region of the color-period plane occupied by close binary stars (Messina et al. in preparation).\\ 
We found that the gravitational forces between the two components of the AT Mic system are about a factor $\sim$7.3$\times10^6$ more intense than those between AU Mic and AT Mic. Therefore, there is the possibility that such intense gravitational forces between AT Mic
 A and B may have enhanced their primordial disc dispersal, shortening the lifetime and, consequently,  the duration of the disk locking phase. On the contrary, the primordial disc of the distant AU Mic ($\sim$46200\,AU), being subjected to a negligible external gravitational perturbation, has evolved with the typical time scale of 4--6 Myr, but rarely longer than about  10 Myr (Ribas et al. 2014; Ingleby et al. 2014), allowing the central star to start its rotation spin up later with respect to AT Mic.

 This possible scenario has been proposed for \rm  the triple BD$-$21\,1074, where a distance of about 16\,AU between the components of the close binary in the system is enough to have significant gravitational effects  (Messina et al. 2014), whereas in the triple TYC\,9300-0891-1AB/TYC\,9300-0525-1, where the distance between the A and B components of TYC9300-0891-1AB is about $\sim$160\,AU,  the effects are negligible (Messina et al. 2016a). 
Also the $\sim$40\% difference between the rotation periods of the \rm components A and B  of AT Mic is very large, considering that they have all 
stellar parameters, age, mass, radius, metallicity  about equal. Such a difference may arise from different initial rotation periods.  In fact, we see in  Fig.\,\ref{distri} that it is comparable to the 3$\sigma$ dispersion observed among similar color single stars. \rm However, another interesting possibility may reside in a different magnetic activity level of the two components.
If one component has a flare activity significantly larger than the other component, then this may lead to a significantly larger loss of angular momentum due to either mass loss or magnetic braking. In this hypothesis, we expect the slower rotating component of AT Mic to be more active (and with an higher rate of flaring) than the faster rotating component of AT Mic. We have attempted to identify if this is the case by investigating on the rate of flare occurrence on the two components, but, as discussed in Sect.\,6, despite we find some evidence that flares are associated more with A component, the quality and amount of our data prevent us to consider these results robust enough. 
To address this interesting issue we need to get resolved time series photometry of both components to check which component is responsible for the flares.\\

\section{Conclusions}
We have analysed the rotational properties of the triple stellar system AU Mic + AT Mic A\&B in the young   25$\pm$3-Myr \rm 
$\beta$ Pictoris association. We have measured the photometric rotation periods P = 1.19\,d and P =  0.78\,d of the AT Mic components, although we could not establish to which components the periods refer. \rm We find that AU Mic is sufficiently distant ($\sim$ 46200\,AU) from AT Mic to have evolved rotationally as a single star
with a primordial disc life time typical of a M1V star. Therefore, its rotation period P = 4.85\,d fits well into the distribution of rotation periods of single members of the $\beta$ Pictoris association. On the contrary,  for the A and B components of AT Mic we propose a scenario according to which they \rm are sufficiently close ($\sim$27\,AU) to have gravitationally perturbed the respective primordial discs, enhancing their dispersion and, consequently,
shortening both the disc life time and the star-disc locking duration.  In this scenario, \rm  AT Mic A and B started to spin up, owing to radius contraction, at earlier epochs with respect to AU Mic, reaching a rotation rate about a factor 5 faster than AU Mic.\\
Interestingly, we find that the A and B components of AT Mic, although substantially equal (same mass, age, metallicity), have a significant ($\sim$40\%) difference between their rotation periods.  This difference may arise from different initial rotation periods. However, we also note that \rm  this system has a high rate of flaring, and we have some
hints that only one component may be the flare star. If this hypothesis would be correct, then the different flaring activity
may be the additional parameter responsible for the different rotational evolution of the two components. In fact, a prominent flare activity can enhance the mass loss rate and, consequently, the angular momentum loss, as well, it can significantly modify the topology of the external magnetic fields making more or less efficient the stellar magnetic braking. 
Certainly, spatially resolved photometry can allow us to see if one or both components have flares and to address this possible
additional cause of the rotation period spread observed among the M-type stars. \\

{\it Acknowledgements}. Research on stellar activity at INAF- Catania Astrophysical Observatory is supported by MIUR  (Ministero dell'Istruzione, dell'Universit\`a e della Ricerca).   This paper makes use of data from the first public release of the WASP data (Butters et al. 2010) as provided by the WASP consortium and services at the NASA Exoplanet Archive, which is operated by the California Institute of Technology, under contract with the National Aeronautics and Space Administration under the Exoplanet Exploration Program. This research has made use of the Simbad database, operated at CDS (Strasbourg, France). We thank the anonymous Referee for useful comments that allowed us to improve the paper quality.

{}
\end{document}